\documentclass[aps,prl,twocolumn,superscriptaddress,groupedaddress]{revtex4}
\usepackage{graphicx}
\usepackage{dcolumn}
\usepackage{bm}
\usepackage{amssymb}

\newcommand{{\edf}}{{\sc edf}}
\newcommand{{\dft}}{{\sc dft}}
\newcommand{{\unedfone}}{\mbox{\sc unedf1}}
\newcommand{\hfbtho}{{\sc hfbtho}}

\begin{document}

\hspace{5.2in} \mbox{LA-UR-19-20003}

\title{Propagation of Statistical Uncertainties of Skyrme Mass Models to
Simulations of $r$-Process Nucleosynthesis}

\author{T.M. Sprouse}
\affiliation{Department of Physics, University of Notre Dame, Notre Dame,
IN 46556, USA}

\author{R. Navarro Perez}
\affiliation{Department of Physics, San Diego State University, San Diego,
CA 02182, USA}

\author{R. Surman}
\affiliation{Department of Physics, University of Notre Dame, Notre Dame,
IN 46556, USA}

\author{M.R. Mumpower}
\affiliation{Theoretical Division, Los Alamos National Laboratory, Los Alamos,
NM 87545, USA}

\author{G.C. McLaughlin}
\affiliation{Department of Physics, North Carolina State University,
Raleigh, NC}

\author{N. Schunck}
\affiliation{Nuclear and Chemical Science Division, Lawrence Livermore National
Laboratory, Livermore, California 94551, USA}

\date{\today}

\begin{abstract}
Uncertainties in nuclear models have a major impact on simulations that aim at
understanding the origin of heavy elements in the universe through the rapid
neutron capture process ($r$ process) of nucleosynthesis. Within the
framework of the nuclear density functional theory, we use results of Bayesian
statistical analysis to propagate uncertainties in the parameters of energy
density functionals to the predicted $r$-process abundance pattern, by way not
only of the nuclear masses but also through the influence of the masses on
$\beta$-decay and neutron capture rates. We additionally make the first
identifications of specific parameters of Skyrme-like energy density
functionals which are correlated with particular aspects of the $r$-process
abundance pattern. While previous studies have explored the reduction in
the abundance pattern uncertainties due to anticipated new measurements of
neutron-rich nuclei, here we point out that an even larger reduction will
occur when these new measurements are used to reduce the uncertainty
of model predictions of masses, which are then propagated through to the
abundance pattern. We make a quantitative prediction for how large this
reduction will be.
\end{abstract}

\maketitle

The heaviest elements owe their origins to rapid neutron capture, or
$r$-process, nucleosynthesis. In the $r$-process, heavy elements are
built up via a sequence of rapid neutron captures and $\beta$-decays
that populate nuclei far to the neutron-rich side of stability
\cite{Burbidge+57,Cameron57}. The astrophysical source of the intense
neutron flux was initially suspected to be within core-collapse supernovae
\cite{Meyer+92,Woosley+94}, though decades of careful study have shown the
required conditions are unlikely to be obtained in this environment
\cite{Arcones+07,Fischer+10,Hudepohl+10,Roberts+12}. Recent evidence
\cite{Hirai+2015,Ji+2016}, including the discovery of
GW170817/GRB170817a/SSS17a \cite{AbbottGW170817,Cowperthwaite2017},
increasingly points to neutron star mergers as the likely $r$-process site.
However, many open questions remain. For example, what specific environments
within neutron star merger events are responsible for $r$-process production,
and what are their properties? Can neutron star mergers account for all
galactic $r$-process production, or are there additional astrophysical sites?

The $r$-process astrophysical conditions could in principle be identified by
comparing simulations of abundance patterns of elements and observations in
the solar system and in old stars. However, analysis of individual
environments is complicated by large uncertainties in the astrophysics
and nuclear physics \cite{Arnould+07}. Here we consider the latter.
Simulations of the $r$-process are dependent upon nuclear data, including
masses, neutron capture rates, and $\beta$-decay and fission properties, for
thousands of neutron-rich nuclei \cite{Mumpower+16}. In spite of a concerted
effort at radioactive beam facilities worldwide to measure these properties
directly or indirectly, the vast majority of them are as of yet inaccessible
and we must rely on theoretical estimates.

Nuclear density functional theory ({\dft}) is currently the only approach that
can provide all of these properties in a consistent yet microscopic framework
\cite{bender2003}. Most energy density functionals ({\edf})
are typically characterized by approximately a dozen
parameters that are fitted on a small set of nuclear properties. The choices
made in selecting the form of the {\edf} and the set of experimental data to
fit its parameters lead to both systematic and statistical uncertainties that
have an impact on all applications \cite{schunck2015-c}.

Ideally, one would like to consider simultaneously all sources of uncertainties
(systematic, statistical and numerical) and propagate them to all observables
(separation energies, $\alpha$-, $\beta$- and $\gamma$-decay rates, fission
rates, neutron capture rates) relevant to astrophysical simulations. Such an
approach is currently not feasible, partly because of its formidable
computational cost, partly because there are still gaps in our understanding
of, e.g., $\alpha$-decay, neutron capture or fission. However, we can exploit
recent work in determining estimates of theoretical uncertainties to quantify
the variations in simulated $r$-process abundances that result from nuclear
mass uncertainties alone. Past work in this area has either considered
abundance pattern comparisons between distinct mass models,
e.g.~\cite{Martin+2016}, or ranges of patterns that result from random,
uncorrelated mass variations \cite{Mumpower+16,Surman+16}.

In this work, we perform the first rigorous propagation of statistical
uncertainties of nuclear mass models based on {\dft}. We generate fifty
different {\edf}s by sampling the Bayesian posterior distribution of the
{\unedfone} {\edf}. For each sample, we compute a full nuclear chart and
update neutron capture rates and $\beta$-decay properties to be consistent
with each table. We implement these sets of nuclear data in $r$-process
simulations to place ``error bars'' due to nuclear masses on $r$-process
abundances and to identify correlations between theoretical model parameters
and abundance pattern features. Such correlations could possibly lead to
additional constraints on $r$-process conditions or, e.g., the {\unedfone}
parameters themselves. Finally, we provide a quantitative estimate of the
improvements to $r$-process pattern uncertainties expected from anticipated
mass measurements at current and upcoming facilities and concurrent
advancements in theoretical models.

We begin by computing atomic mass tables within the nuclear {\dft} approach
to nuclear structure with Skyrme {\edf}s. Our starting point is the
{\unedfone} parametrization, in which the coupling constants were optimized
globally on select experimental nuclear masses, radii, deformations and
excitation energies of fission isomers in the actinides
\cite{Kortelainen:2011ft}. While the r.m.s. deviation on nuclear binding
energies of {\unedfone} is only 1.8 MeV, it goes down to 0.45 MeV for
2-neutron separation energies. Bayesian inference methods were later used to
compute the posterior distribution of the {\unedfone}
parameters~\cite{Higdon:2014tva} and propagate theoretical statistical
uncertainties in predictions of nuclear masses, two-neutron drip line,
and fission barriers~\cite{McDonnell:2015sja}. Here, we sample the same
posterior distribution within the $90\%$ confidence region to generate
fifty different parameter sets for the Skyrme {\edf}.

For each sample, we compute the nuclear ground-state binding energy
of all even-even nuclei from Hydrogen to $Z=120$ by solving the
Hartree-Fock-Bogoliubov (HFB) equation. The limits of nuclear
stability (proton and neutron drip lines) are reached when the value of the
two neutron (proton) separation energy becomes negative. Compared to
alternative options based, e.g., on the value of the Fermi energy, this
criterion offers the advantage of being model-independent since binding
energies are true observables. With this criterion, each mass table contains
approximately 2,000 even-even nuclei. For each even-even nucleus, the
ground-state is determined by exploring locally the potential energy
surface of the nucleus for a range of eleven axial quadrupole deformations
$\beta_2$ between -0.5 and +0.5. The configuration with the lowest energy
defines the ground-state. Details of the exploration of the even-even nuclear
landscape with the numerical solver {\hfbtho} can be found in \cite{perez2017}.
With this procedure, computing 50 mass tables requires of the order of
1 million HFB calculations.

Although odd-even and odd-odd binding energies could be computed with the
blocking procedure, see, e.g. \cite{schunck2010}, this would require about
an order of magnitude more HFB calculations. Instead we adopt a standard
approximation for the binding energy of odd nuclei that combines information
about binding energies and HFB pairing gaps in neighboring isotopes/isotones;
see Supplemental Material of \cite{erler2012}. This procedure yields an
excellent approximation of, in particular, one-particle separation
energies.

\begin{figure}[!ht]
\begin{center}
\includegraphics[width=\columnwidth]{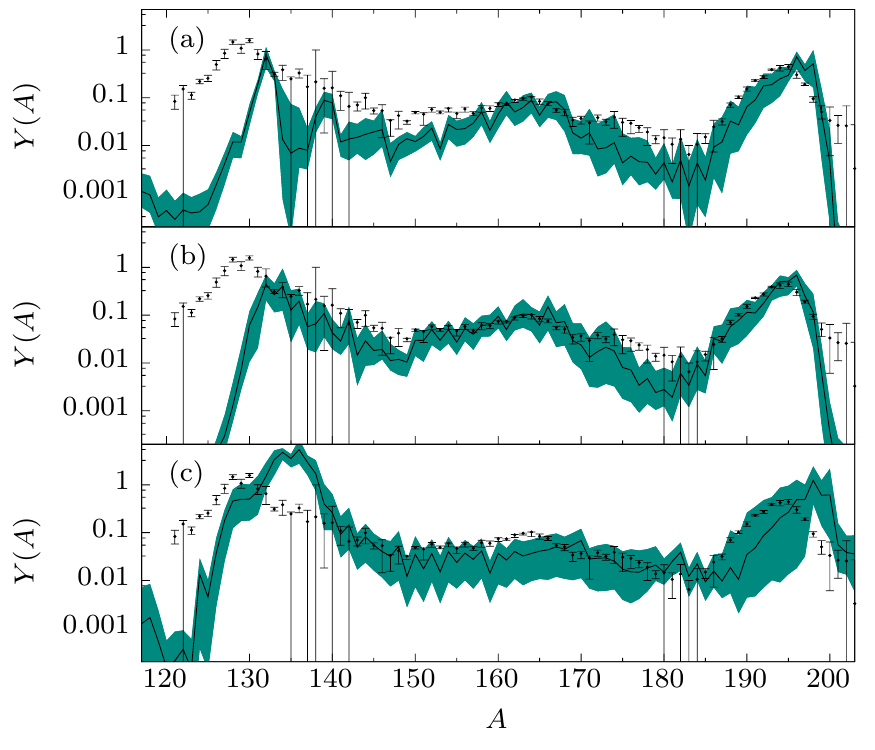}
\caption{\label{fig:YvA} Abundance patterns $Y(A)$ versus $A$ for fifty
$r$-process simulations with astrophysical conditions corresponding to
high-entropy (top panel, (a)), low-entropy (middle panel, (b)), and
fission-recycling (bottom panel, (c)) outflows, as described in the text.
The shaded region shows the full range of abundance patterns produced, and
the black line shows their mean. All patterns are scaled to solar abundances
from \cite{Arnould+07}. }
\end{center}
\end{figure}

For each of the fifty mass tables thus described, we calculate a
self-consistent set of all nuclear data inputs required for
\textit{r-}process calculations. We calculate neutron capture and
neutron-induced fission rates using the Los Alamos Hauser-Feshbach
code CoH \cite{Kawano+16} and $\beta$-decay half-lives with
probabilities for delayed emission of one or more neutrons using the QRPA+HF
framework of \cite{MumpQRPA+HF} and unmodified strength data
from \cite{Moller+97}. We repeat these calculations using the
masses given in the 2016 Atomic Mass Evaluation (AME2016) \cite{wang2017};
where possible, these results are taken to replace those
based on the {\unedfone} mass tables. The decay properties of the Nubase 2016
compilation \cite{NUBASE2016} are further taken to replace any calculated
values based on either AME2016 or {\unedfone} nuclear masses. For all
fissioning nuclei, we use a symmetric, two-particle product distribution.

We implement each of these datasets into the nuclear reaction network code
PRISM \cite{Mumpower+17b,Mumpower+18,Holmbeck+18} to simulate nucleosynthesis
for three distinct types of astrophysical conditions where \textit{r-}process
nucleosynthesis may occur: (1) a supernova-type high-entropy wind, with
entropy $s/k=300$, dynamical timescale $\tau=80$ ms, and electron fraction
$Y_{e}=0.30$, (2) a parameterized merger accretion disk wind with $s/k=30$,
$\tau=80$ ms, and $Y_{e}=0.21$, and (3) fission-recycling outflow from a
neutron star merger \cite{Mendoza+15}. For each simulation, we dynamically
update the evolution of temperature with respect to the release of energy
from nuclear reactions, decays, and fission, with an assumed thermalization
efficiency of $10\%$ for all energy released.

The range in final abundance patterns across these fifty calculations is shown
in Fig.~\ref{fig:YvA} for each set of astrophysical conditions we consider.
In each case, the shaded band represents the propagation of statistical
uncertainties from the {\unedfone} nuclear mass model to the corresponding
$r$-process simulation. The influence of masses on reaction and decay rates
contribute to the width of the band, via mechanisms described in, e.g.,
\cite{Mumpower+16} and references therein. In addition, the location of
the neutron drip line is of key importance for some types of astrophysical
conditions. In the fifty mass tables considered here, the location of the
one-neutron drip line varies by more than ten neutron numbers. Notably, the
band is the widest for the astrophysical conditions in which the $r$-process
path is the closest to the drip line, the fission
recycling example of Fig.~\ref{fig:YvA}(c).

\begin{figure}[!ht]
\begin{center}
\includegraphics[width=\columnwidth]{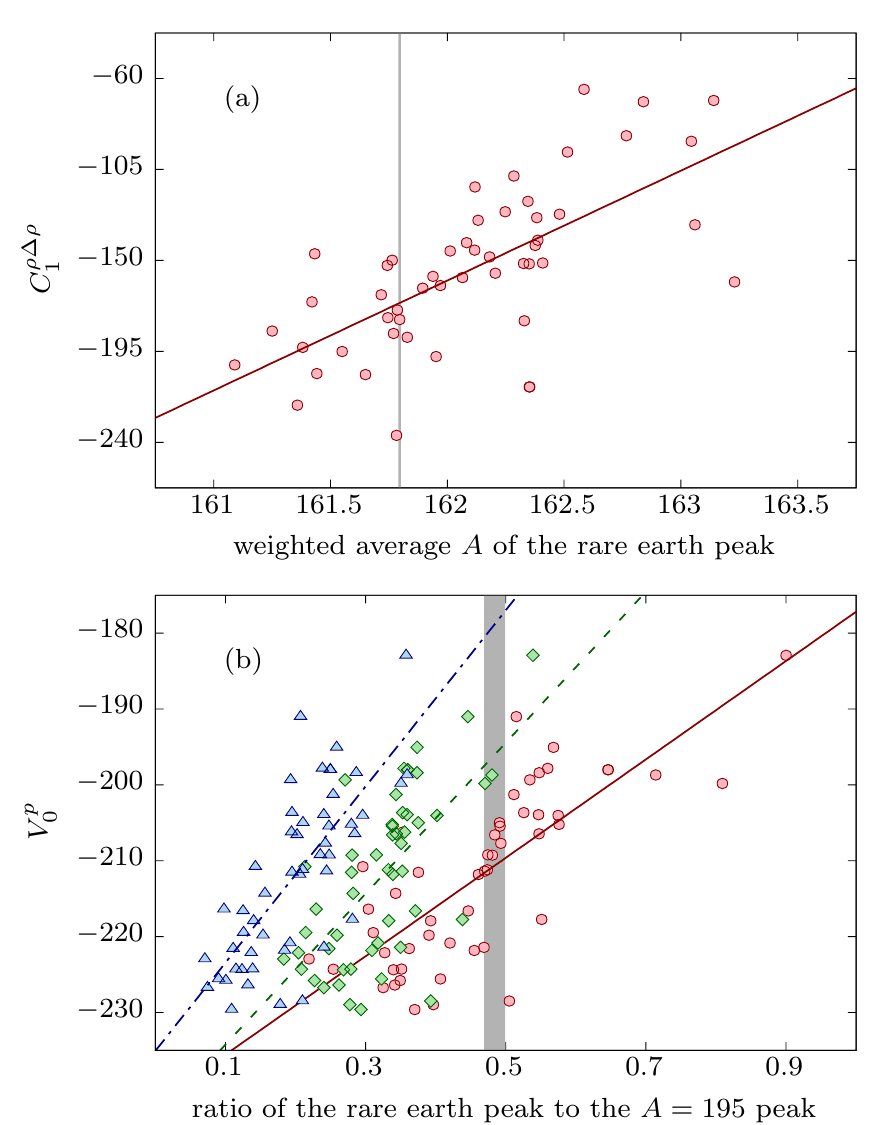}
\caption{\label{fig:corr}
Relationship between $r$-process abundance pattern and {\unedfone} functional
parameters for fifty {\unedfone} mass tables. Panel (a) shows the relationship
between the weighted average mass number $A$ of the rare earth peak and the
isovector surface coupling constant $C_{1}^{\rho \Delta \rho}$ for the
low-entropy wind conditions of Fig.~\ref{fig:YvA}. A linear fit to the
dataset is given by the solid line with correlation coefficient $r=0.68$.
Panel (b) shows the relationship between the proton pairing strength
$V_{0}^{p}$ and the ratio of summed abundances in the rare earth region
to the $A=195$ region for the high-entropy (green diamonds), low-entropy
(red circles), and fission recycling (blue triangles) conditions from
Fig.~\ref{fig:YvA}, with linear fits  given for the high-entropy dataset
by the green dashed line ($r=0.66$), the low-entropy dataset by the  red
solid line ($r=0.76$), and the fission recycling dataset by the blue
dot-dashed line ($r=0.75$). The gray shaded region in each figure indicates
the range of values in each metric admitted by the solar abundances
of  \cite{Arnould+07} and \cite{Sneden+08}.}
\end{center}
\end{figure}

With the wealth of data available from these $r$-process simulations, 
we can search for correlations between {\unedfone}
functional parameters and the formation of abundance pattern features. Here
we demonstrate how such analyses might proceed.

Several of the {\unedfone} functional parameters are poorly constrained by
data near stability. One such parameter is the isovector surface coupling
constant $C_{1}^{\rho \Delta \rho}$, with {\unedfone} value
$-145.382\pm 52.169$; see Table II in \cite{Kortelainen:2011ft}. For a
low-entropy hot wind $r$-process environment, this parameter is correlated
with the formation of the rare earth peak, the small feature around
$A\sim 160$ in the solar $r$-process isotopic pattern. The top panel of
Fig.~\ref{fig:corr} shows $C_{1}^{\rho \Delta \rho}$ versus the
abundance-weighted average $A$ of the rare earth peak for the $r$-process
simulations from the middle panel of Fig.~\ref{fig:YvA}. The placement of
the rare earth peak is calculated from the solar $r$-process abundances
of \cite{Arnould+07} and \cite{Sneden+08} and is given by the shaded vertical
band. Correlations between $C_{1}^{\rho \Delta \rho}$ and rare earth peak
placement are weaker for other types of $r$-process environments.
The $r$-process path in the high-entropy wind case is not so neutron-rich and
thus not as sensitive to $C_{1}^{\rho \Delta \rho}$. The fission recycling
example has a distinct rare earth peak formation
mechanism \cite{Mumpower+17} that is not particularly active with
the {\unedfone} masses, resulting in a comparatively weaker correlation.
However, recent studies \cite{mclaughlin+2005,surman+2006,Just+2015,
Martin+2015,Wanajo+2014,Siegel+2017} favor $r$-process conditions that
are most similar to those of our low-entropy wind where this
correlation is strongest,  suggesting that the $r$-process abundance pattern
may provide an important additional constraint on the value
of $C_{1}^{\rho \Delta \rho}.$

In all of the astrophysical environments considered, we found the proton
pairing strength $V_{0}^{p}$ to be correlated with the ratio between the
summed abundances of the rare earth and $A\sim 195$ peak regions, as
illustrated in the bottom panel of Fig.~\ref{fig:corr}, where the solar
values are given by the shaded band. The correlations in each case
are distinct, with different astrophysical conditions picking out different
preferred values of $V_{0}^{p}$. Only the least negative values of $V_0^p$
considered reproduce solar values for the high-entropy conditions, while
values of $V_0^p$ that tend towards the center of the distribution reproduce
solar values for the low-entropy conditions. Within the range of values we
consider, the fission recycling conditions fail to reproduce
solar values, with the correlation suggesting an even more negative value of
$V_{0}^{p}.$ Thus, if $V_{0}^{p}$ could be more tightly constrained, the
simulated ratio of the rare earth and $A\sim 195$ peak regions could be used
as a diagnostic of $r$-process conditions.

\begin{figure}[!ht]
\begin{center}
\includegraphics[width=\columnwidth]{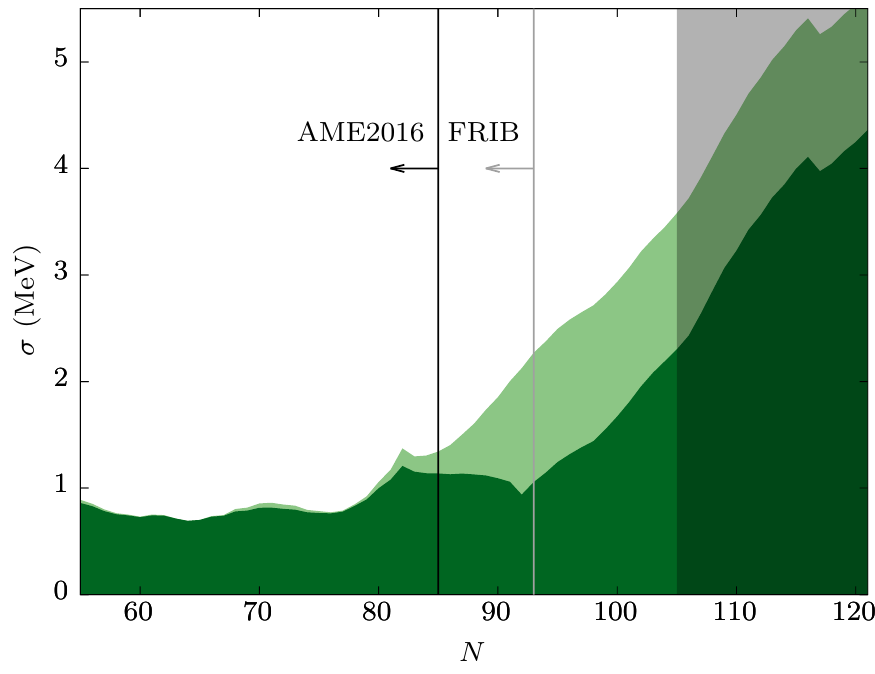}
\caption{\label{fig:sn}  Variations $\sigma$ among our set of 50
{\unedfone} mass tables (light green shaded region) and our set
of simulated mass tables (dark green shaded region), each with
respect to the nominal {\unedfone} masses, for the tin isotopes.
The AME2016 range of known masses \cite{wang2017}
and anticipated FRIB reach are indicated, respectively, by black
and gray solid lines. The vertical darkened band
indicates the range in location for the one-neutron dripline.}
\end{center}
\end{figure}

Measurements of the masses of increasingly neutron-rich nuclei are the focus
of a number of experimental efforts worldwide, for example at the Canadian
Penning Trap at CARIBU \cite{Hirsh+2016,Orford+2018},
JYFLTRAP at Jyv{\"a}skyl{\"a} \cite{Kankainen+2013,Vilen+2018}, ISOLTRAP at
CERN \cite{Lunney+2017}, TITAN at TRIUMF \cite{Lascar+2017}, and storage
rings at GSI in Germany, IMP in China, and RIKEN in Japan \cite{Zhang+2016}.
Next-generation radioactive ion facilities, such as the Facility for Rare
Isotope Beams (FRIB), will
have unprecedented access to isotopes far from stability \cite{FRIB}. New
mass measurements improve the reliability of $r$-process simulations in
two ways: directly, by dramatically reducing the uncertainty in the masses
of newly measured nuclei, and indirectly, by enabling improvements to mass
modeling. Theoretical mass models are all calibrated to known
data, so known masses tend to be well reproduced by theory. Outside the known
region, theoretical predictions tend to diverge. The variations among our
fifty {\unedfone} mass tables, shown for the tin isotopes in Fig.~\ref{fig:sn},
clearly demonstrate this behavior. For this element, {\unedfone} fits known
masses to about $\sigma_{rms} \sim 1$ MeV, and variations increase sharply
past the $N=82$ closed shell. Additional measurements increase the available
data with which to constrain theory and thus hold the potential to reduce
uncertainties outside the measured region. We simulate this effect by
generating an adjusted set of fifty mass tables, in which the variations from
the mean are reduced to match the experimentally-known region for nuclei within
the FRIB range and increase with the same slope outside this range.
The variations of the simulated set of tables are shown in the dark shaded
region of Fig.~\ref{fig:sn}.

Our two sets of {\unedfone} mass tables can be used to quantify the reductions
in $r$-process abundance pattern uncertainties that have already been achieved
by measurements to date and that are anticipated from future mass measurements.
We rerun the example $r$-process simulations from Fig.~\ref{fig:YvA} using
three different sets of nuclear data. The first set is a theory-only set,
with all quantities derived exclusively from our fifty {\unedfone} tables.
The second set is that used in Fig.~\ref{fig:YvA}, where experimental nuclear
data is additionally incorporated. The third set is constructed to mimic the
influence of anticipated mass measurements. Experimental values or values
derived from the nominal {\unedfone} mass tables are held fixed
for all nuclei within the FRIB reach;
elsewhere we use theory values derived from
our set of fifty simulated {\unedfone} mass tables.

\begin{figure}[!ht]
\begin{center}
\includegraphics[width=\columnwidth]{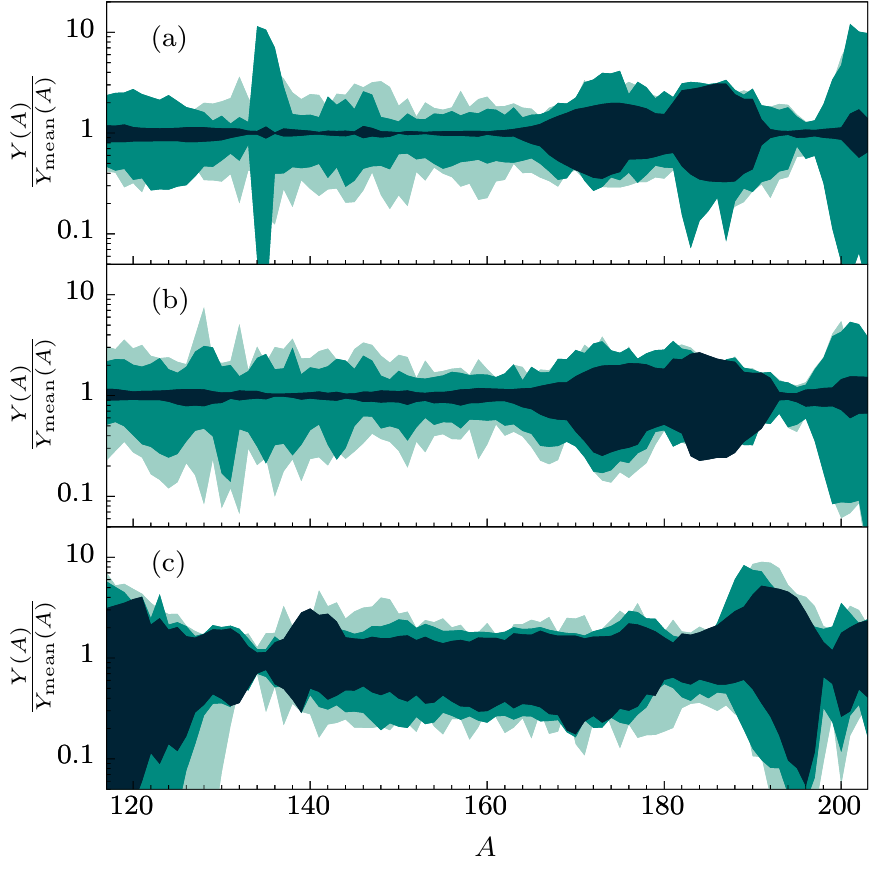}
\caption{\label{fig:yymean} Ratios of the abundances $Y(A)$ to the mean
abundance $Y_{\text{mean}}(A)$ for the set of fifty simulations with the
example high entropy wind (top panel, (a)), low entropy wind (middle panel,
(b)), and fission recycling outflow (bottom panel, (c)) astrophysical
conditions, as in Fig.~\ref{fig:YvA}. The light shaded band shows
theory-only calculations, the medium shaded band implements AME2016 masses
and NUBASE2016 decay properties where available, and the dark shaded band
additionally includes the simulated mass tables described in the text.}
\end{center}
\end{figure}

Fig.~\ref{fig:yymean} shows the abundance pattern variations normalized
by the mean for the three astrophysical trajectories in Fig.~\ref{fig:YvA},
each calculated with the three data sets described above. For the high entropy
wind, Fig.~\ref{fig:yymean}(a), many of the relevant
nuclear properties have already been measured, so there is significant
improvement realized between the theory-only (lightest shaded band)
calculations and those that include current experimental values (medium shaded
band). Looking forward to FRIB, the majority of nuclei along the equilibrium
$r$-process path in the $N=82$ and rare earth regions will be within reach.
Thus systematic measurement campaigns at FRIB have the potential to essentially
remove mass as a source of uncertainty in simulated $r$-process abundances
below $A\sim170$ for high entropy winds.

However, in the currently-favored potential $r-$process astrophysical site of
neutron star/neutron star-black hole mergers, the environments are likely
lower entropy, $s/k\sim 5-50$, and more neutron-rich, similar to the
conditions used for the middle and bottom panels of Figs.~\ref{fig:YvA}
and \ref{fig:yymean}. The $r$-process equilibrium paths are farther from
stability in these cases, thus the current reach of experimental data results
in more modest improvements, as indicated when comparing the light- and
medium-shaded bands in Fig.~\ref{fig:yymean}(b) and (c). Prospects for the
future, however, are encouraging. For the low-entropy wind example
of Fig.~\ref{fig:yymean}(b), FRIB can reach the majority of the key nuclei
and the remaining uncertainty band should be similar to the high-entropy
wind case. In particular the excellent precision anticipated for abundances
$140<A<170$ can facilitate the use of the rare earth peak as a
key $r$-process diagnostic \cite{Mumpower+12,Orford+2018}.

For the fission recycling example, uncertainties in the location of the drip
line and in the fission properties of heavy nuclei near the drip line dominate
the uncertainty bands. Even with FRIB at full power these uncertainties are
unlikely to be resolved with direct measurements. Here nuclear theory will
play a critical role. The simulated improvements to theory anticipated in
our approach do result in a narrowing of the uncertainty band, as seen in a
comparison between the medium- and dark-shaded bands
of Fig.~\ref{fig:yymean}(c). Further potential improvements to nuclear
EDF theory, e.g.~\cite{Neufcourt+2018}, and its full application to the
problem of fission, e.g.~\cite{Giuliani+2018,Bulgac+18}, are not captured in
our approach. Therefore, there remains the possibility for more significant
improvements to the uncertainty band associated with fission-recycling
conditions with concurrent advances in experiment and theory.

The origins of the heaviest elements have remained mysterious for decades.
Thanks to concerted efforts in astrophysical modeling, spectroscopic
observations, neutrino and nuclear experiment and theory, and, now,
gravitational wave astronomy, a detailed understanding of $r$-process
nucleosynthesis finally seems within reach. Still, further advances are
needed in each of these areas. Here we have highlighted how careful
quantification of nuclear physics uncertainties has the potential to
provide crucial insight into $r$-process astrophysical conditions and
the nuclear models themselves.

\begin{acknowledgements}
This work was supported in part by the U.S. Department of Energy under
grant numbers DE-SC0013039 (RS), DE-FG02-95-ER40934 (RS),
DE-FG02-02ER41216 (GCM), and DE-FG02-93ER-40756 (RNP), and the SciDAC
collaborations TEAMS DE-SC0018232 (TS, RS) and NUCLEI DE-SC0018223 (RNP, NS).
Part of this work was carried out under the auspices of the National Nuclear
Security Administration of the U.S. Department of Energy at Los Alamos
National Laboratory under Contract No. DE-AC52-06NA25396 (MM); under the
auspices of the U.S.\ Department of Energy by Lawrence Livermore National
Laboratory under Contract DE-AC52-07NA27344 (NS); under the FIRE topical
collaboration in nuclear theory funded by the U.S.\ Department of Energy
under contract DE-AC52-07NA27344 (RS, MM, GCM). Computing support for this
work came from the Lawrence Livermore National Laboratory (LLNL) Institutional
Computing Grand Challenge program.
\end{acknowledgements}

\bibliographystyle{unsrt}
\bibliography{}

\begin{thebibliography}{10}

\bibitem{Burbidge+57}
E.~M. {Burbidge}, G.~R. {Burbidge}, W.~A. {Fowler}, and F.~{Hoyle}.
\newblock {Synthesis of the Elements in Stars}.
\newblock {\em Reviews of Modern Physics}, 29:547--650, 1957.

\bibitem{Cameron57}
A.~G.~W. {Cameron}.
\newblock {Nuclear Reactions in Stars and Nucleogenesis}.
\newblock {\em Chalk River Reports}, CRL-41, 1957.

\bibitem{Meyer+92}
B.~S. {Meyer}, G.~J. {Mathews}, W.~M. {Howard}, S.~E. {Woosley}, and R.~D.
  {Hoffman}.
\newblock {R-process nucleosynthesis in the high-entropy supernova bubble}.
\newblock {\em Astrophys.\ J.}, 399:656--664, November 1992.

\bibitem{Woosley+94}
S.~E. {Woosley}, J.~R. {Wilson}, G.~J. {Mathews}, R.~D. {Hoffman}, and B.~S.
  {Meyer}.
\newblock {The r-process and neutrino-heated supernova ejecta}.
\newblock {\em Astrophys.\ J.}, 433:229--246, September 1994.

\bibitem{Arcones+07}
A.~{Arcones}, H.-T. {Janka}, and L.~{Scheck}.
\newblock {Nucleosynthesis-relevant conditions in neutrino-driven supernova
  outflows. I. Spherically symmetric hydrodynamic simulations}.
\newblock {\em Astron.\ Astrophys.}, 467:1227--1248, June 2007.

\bibitem{Fischer+10}
T.~{Fischer}, S.~C. {Whitehouse}, A.~{Mezzacappa}, F.-K. {Thielemann}, and
  M.~{Liebend{\"o}rfer}.
\newblock {Protoneutron star evolution and the neutrino-driven wind in general
  relativistic neutrino radiation hydrodynamics simulations}.
\newblock {\em Astron.\ Astrophys.}, 517:A80, July 2010.

\bibitem{Hudepohl+10}
L.~{H{\"u}depohl}, B.~{M{\"u}ller}, H.-T. {Janka}, A.~{Marek}, and G.~G.
  {Raffelt}.
\newblock {Neutrino Signal of Electron-Capture Supernovae from Core Collapse to
  Cooling}.
\newblock {\em Physical Review Letters}, 104(25):251101, June 2010.

\bibitem{Roberts+12}
L.~F. {Roberts}, S.~{Reddy}, and G.~{Shen}.
\newblock {Medium modification of the charged-current neutrino opacity and its
  implications}.
\newblock {\em Phys.\ Rev.\ C}, 86(6):065803, December 2012.

\bibitem{Hirai+2015}
Y.~{Hirai}, Y.~{Ishimaru}, T.~R. {Saitoh}, M.~S. {Fujii}, J.~{Hidaka}, and
  T.~{Kajino}.
\newblock {Enrichment of r-process Elements in Dwarf Spheroidal Galaxies in
  Chemo-dynamical Evolution Model}.
\newblock {\em Astrophys.\ J.}, 814:41, November 2015.

\bibitem{Ji+2016}
A.~P. {Ji}, A.~{Frebel}, A.~{Chiti}, and J.~D. {Simon}.
\newblock {R-process enrichment from a single event in an ancient dwarf
  galaxy}.
\newblock {\em Nature}, 531:610--613, March 2016.

\bibitem{AbbottGW170817}
B.~P. {Abbott \textit{et al.}}
\newblock Gw170817: Observation of gravitational waves from a binary neutron
  star inspiral.
\newblock {\em Phys.\ Rev.\ Lett.}, 119:161101, Oct 2017.

\bibitem{Cowperthwaite2017}
P.~S. {Cowperthwaite \textit{et al.}}
\newblock The electromagnetic counterpart of the binary neutron star merger
  ligo/virgo gw170817. ii. uv, optical, and near-infrared light curves and
  comparison to kilonova models.
\newblock {\em The Astrophysical Journal Letters}, 848(2):L17, 2017.

\bibitem{Arnould+07}
M.~{Arnould}, S.~{Goriely}, and K.~{Takahashi}.
\newblock {The r-process of stellar nucleosynthesis: Astrophysics and nuclear
  physics achievements and mysteries}.
\newblock {\em Physics Reports}, 450:97--213, September 2007.

\bibitem{Mumpower+16}
M.~R. {Mumpower}, R.~{Surman}, G.~C. {McLaughlin}, and A.~{Aprahamian}.
\newblock {\em Progress in Particle and Nuclear Physics}, 86:86--126, January
  2016.

\bibitem{bender2003}
Michael Bender, Paul-Henri Heenen, and Paul-Gerhard Reinhard.
\newblock {Self-consistent} mean-field models for nuclear structure.
\newblock {\em Rev. Mod. Phys.}, 75(1):121, 2003.

\bibitem{schunck2015-c}
N.~Schunck, J.~D. McDonnell, D.~Higdon, J.~Sarich, and S.~M. Wild.
\newblock {Uncertainty} quantification and propagation in nuclear density
  functional theory.
\newblock {\em Eur. Phys. J. A}, 51(12):1, 2015.

\bibitem{Martin+2016}
D.~{Martin}, A.~{Arcones}, W.~{Nazarewicz}, and E.~{Olsen}.
\newblock {Impact of Nuclear Mass Uncertainties on the r Process}.
\newblock {\em Physical Review Letters}, 116(12):121101, March 2016.

\bibitem{Surman+16}
R.~{Surman}, M.~{Mumpower}, and A.~{Aprahamian}.
\newblock {Uncorrelated Nuclear Mass Uncertainties and r-process Abundance
  Predictions}.
\newblock {\em Acta Physica Polonica B}, 47:673, 2016.

\bibitem{Kortelainen:2011ft}
M.~Kortelainen, J.~McDonnell, W.~Nazarewicz, P.~G. Reinhard, J.~Sarich,
  N.~Schunck, M.~V. Stoitsov, and S.~M. Wild.
\newblock {Nuclear energy density optimization: Large deformations}.
\newblock {\em Phys. Rev.}, C85:024304, 2012.

\bibitem{Higdon:2014tva}
Dave Higdon, Jordan~D. McDonnell, Nicolas Schunck, Jason Sarich, and Stefan~M
  Wild.
\newblock {A Bayesian Approach for Parameter Estimation and Prediction using a
  Computationally Intensive Model}.
\newblock {\em J. Phys.}, G42(3):034009, 2015.

\bibitem{McDonnell:2015sja}
J.~D. McDonnell, N.~Schunck, D.~Higdon, J.~Sarich, S.~M. Wild, and
  W.~Nazarewicz.
\newblock {Uncertainty Quantification for Nuclear Density Functional Theory and
  Information Content of New Measurements}.
\newblock {\em Phys. Rev. Lett.}, 114:122501, 2015.

\bibitem{perez2017}
R.~Navarro Perez, N.~Schunck, R.~D. Lasseri, C.~Zhang, and J.~Sarich.
\newblock {Axially} deformed solution of the
  {Skyrme--Hartree--Fock--Bogolyubov} equations using the transformed harmonic
  oscillator basis {(III)} {{HFBTHO}} (v3.00): {A} new version of the program.
\newblock {\em Comput. Phys. Commun.}, 220(Supplement C):363, 2017.

\bibitem{schunck2010}
N.~Schunck, J.~Dobaczewski, J.~McDonnell, J.~Moré, W.~Nazarewicz, J.~Sarich,
  and M.~V. Stoitsov.
\newblock {One-quasiparticle} states in the nuclear energy density functional
  theory.
\newblock {\em Phys. Rev. C}, 81(2):024316, 2010.

\bibitem{erler2012}
Jochen Erler, Noah Birge, Markus Kortelainen, Witold Nazarewicz, Erik Olsen,
  Alexander~M. Perhac, and Mario Stoitsov.
\newblock {The} limits of the nuclear landscape.
\newblock {\em Nature}, 486(7404):509, 2012.

\bibitem{Kawano+16}
T.~{Kawano}, R.~{Capote}, S.~{Hilaire}, and P.~{Chau Huu-Tai}.
\newblock {\em Physical Review C}, 94(1):014612, July 2016.

\bibitem{MumpQRPA+HF}
M.~R. Mumpower, T.~Kawano, and P.~M\"oller.
\newblock Neutron-$\ensuremath{\gamma}$ competition for
  $\ensuremath{\beta}$-delayed neutron emission.
\newblock {\em Phys. Rev. C}, 94:064317, Dec 2016.

\bibitem{Moller+97}
P.~{M{\"o}ller}, J.~R. {Nix}, and K.-L. {Kratz}.
\newblock {\em Atomic Data and Nuclear Data Tables}, 66:131, 1997.

\bibitem{wang2017}
Meng Wang, G.~Audi, F.~G. Kondev, W.~J. Huang, S.~Naimi, and Xing Xu.
\newblock {The} {AME2016} atomic mass evaluation {(II).} {Tables}, graphs and
  references.
\newblock {\em Chinese Phys. C}, 41(3):030003, 2017.

\bibitem{NUBASE2016}
G.~Audi et~al.
\newblock The nubase2016 evaluation of nuclear properties.
\newblock {\em Chinese Physics C}, 41(3):030001, March 2017.

\bibitem{Mumpower+17b}
M.~R. Mumpower, T.~Kawano, J.~L. Ullmann, M.~Krtička, and T.~M. Sprouse.
\newblock Estimation of {M} 1 scissors mode strength for deformed nuclei in the
  medium- to heavy-mass region by statistical {Hauser}-{Feshbach} model
  calculations.
\newblock {\em Physical Review C}, 96(2):024612, August 2017.

\bibitem{Mumpower+18}
M.~R. Mumpower, T.~Kawano, T.~M. Sprouse, N.~Vassh, E.~M. Holmbeck, R.~Surman,
  and P.~Möller.
\newblock $\beta$-delayed {Fission} in r-process {Nucleosynthesis}.
\newblock {\em The Astrophysical Journal}, 869(1):14, December 2018.

\bibitem{Holmbeck+18}
Erika~M. Holmbeck, Trevor~M. Sprouse, Matthew~R. Mumpower, Nicole Vassh,
  Rebecca Surman, Timothy~C. Beers, and Toshihiko Kawano.
\newblock Actinide {Production} in the {Neutron}-rich {Ejecta} of a {Neutron}
  {Star} {Merger}.
\newblock {\em The Astrophysical Journal}, 870(1):23, December 2018.

\bibitem{Mendoza+15}
Joel de~Jes\'us Mendoza-Temis, Meng-Ru Wu, Karlheinz Langanke, Gabriel
  Mart\'{\i}nez-Pinedo, Andreas Bauswein, and Hans-Thomas Janka.
\newblock Nuclear robustness of the $r$ process in neutron-star mergers.
\newblock {\em Phys. Rev. C}, 92:055805, Nov 2015.

\bibitem{Sneden+08}
C.~{Sneden}, J.~J. {Cowan}, and R.~{Gallino}.
\newblock {Neutron-Capture Elements in the Early Galaxy}.
\newblock {\em Annual Review of Astronomy and Astrophysics}, 46:241--288,
  September 2008.

\bibitem{Mumpower+17}
M.~R. {Mumpower}, G.~C. {McLaughlin}, R.~{Surman}, and A.~W. {Steiner}.
\newblock {Reverse engineering nuclear properties from rare earth abundances in
  the r process}.
\newblock {\em Journal of Physics G Nuclear Physics}, 44(3):034003, March 2017.

\bibitem{mclaughlin+2005}
G.C. McLaughlin and R.~Surman.
\newblock Prospects for obtaining an r process from {Gamma} {Ray} {Burst}
  {Disk} {Winds}.
\newblock {\em Nuclear Physics A}, 758:189--196, July 2005.

\bibitem{surman+2006}
R.~Surman, G.~C. McLaughlin, and W.~R. Hix.
\newblock Nucleosynthesis in the {Outflow} from {Gamma}-{Ray} {Burst}
  {Accretion} {Disks}.
\newblock {\em The Astrophysical Journal}, 643(2):1057--1064, June 2006.

\bibitem{Just+2015}
O.~Just, A.~Bauswein, R.~Ardevol Pulpillo, S.~Goriely, and H.-T. Janka.
\newblock Comprehensive nucleosynthesis analysis for ejecta of compact binary
  mergers.
\newblock {\em Monthly Notices of the Royal Astronomical Society},
  448(1):541--567, March 2015.

\bibitem{Martin+2015}
D.~Martin, A.~Perego, A.~Arcones, F.-K. Thielemann, O.~Korobkin, and
  S.~Rosswog.
\newblock Neutrino-driven winds in the aftermath of a neutron star merger:
  nucleosynthesis and electromagnetic transients.
\newblock {\em The Astrophysical Journal}, 813(1):2, October 2015.

\bibitem{Wanajo+2014}
Shinya Wanajo, Yuichiro Sekiguchi, Nobuya Nishimura, Kenta Kiuchi, Koutarou
  Kyutoku, and Masaru Shibata.
\newblock Production of all the r-process nuclides in the dynamical ejecta of
  neutron star mergers.
\newblock {\em The Astrophysical Journal}, 789(2):L39, June 2014.

\bibitem{Siegel+2017}
Daniel~M. Siegel and Brian~D. Metzger.
\newblock Three-{Dimensional} {General}-{Relativistic} {Magnetohydrodynamic}
  {Simulations} of {Remnant} {Accretion} {Disks} from {Neutron} {Star}
  {Mergers}: {Outflows} and r -{Process} {Nucleosynthesis}.
\newblock {\em Physical Review Letters}, 119(23):231102, December 2017.

\bibitem{Hirsh+2016}
T.~Y. {Hirsh}, N.~{Paul}, M.~{Burkey}, A.~{Aprahamian}, F.~{Buchinger},
  S.~{Caldwell}, J.~A. {Clark}, A.~F. {Levand}, L.~L. {Ying}, S.~T. {Marley},
  G.~E. {Morgan}, A.~{Nystrom}, R.~{Orford}, A.~P. {Galv{\'a}n}, J.~{Rohrer},
  G.~{Savard}, K.~S. {Sharma}, and K.~{Siegl}.
\newblock {First operation and mass separation with the CARIBU MR-TOF}.
\newblock {\em Nuclear Instruments and Methods in Physics Research B},
  376:229--232, June 2016.

\bibitem{Orford+2018}
R.~{Orford}, N.~{Vassh}, J.~A. {Clark}, G.~C. {McLaughlin}, M.~R. {Mumpower},
  G.~{Savard}, R.~{Surman}, A.~{Aprahamian}, F.~{Buchinger}, M.~T. {Burkey},
  D.~A. {Gorelov}, T.~Y. {Hirsh}, J.~W. {Klimes}, G.~E. {Morgan}, A.~{Nystrom},
  and K.~S. {Sharma}.
\newblock {Precision Mass Measurements of Neutron-Rich Neodymium and Samarium
  Isotopes and Their Role in Understanding Rare-Earth Peak Formation}.
\newblock {\em Physical Review Letters}, 120(26):262702, June 2018.

\bibitem{Kankainen+2013}
A.~{Kankainen}, J.~{Hakala}, T.~{Eronen}, D.~{Gorelov}, A.~{Jokinen}, V.~S.
  {Kolhinen}, I.~D. {Moore}, H.~{Penttil{\"a}}, S.~{Rinta-Antila},
  J.~{Rissanen}, A.~{Saastamoinen}, V.~{Sonnenschein}, and J.~{{\"A}yst{\"o}}.
\newblock {Isomeric states close to doubly magic $^{132}$Sn studied with the
  double Penning trap JYFLTRAP}.
\newblock {\em Phys.\ Rev.\ C}, 87(2):024307, February 2013.

\bibitem{Vilen+2018}
M.~{Vilen}, J.~M. {Kelly}, A.~{Kankainen}, M.~{Brodeur}, A.~{Aprahamian},
  L.~{Canete}, T.~{Eronen}, A.~{Jokinen}, T.~{Kuta}, I.~D. {Moore}, M.~R.
  {Mumpower}, D.~A. {Nesterenko}, H.~{Penttil{\"a}}, I.~{Pohjalainen}, W.~S.
  {Porter}, S.~{Rinta-Antila}, R.~{Surman}, A.~{Voss}, and J.~{{\'n}yst{\"o}}.
\newblock {Precision Mass Measurements on Neutron-Rich Rare-Earth Isotopes at
  JYFLTRAP: Reduced Neutron Pairing and Implications for r -Process
  Calculations}.
\newblock {\em Physical Review Letters}, 120(26):262701, June 2018.

\bibitem{Lunney+2017}
D.~{Lunney} and {(on behalf of ISOLTRAP Collaboration}.
\newblock {Extending and refining the nuclear mass surface with ISOLTRAP}.
\newblock {\em Journal of Physics G Nuclear Physics}, 44(6):064008, June 2017.

\bibitem{Lascar+2017}
D.~{Lascar}, R.~{Klawitter}, C.~{Babcock}, E.~{Leistenschneider}, S.~R.
  {Stroberg}, B.~R. {Barquest}, A.~{Finlay}, M.~{Foster}, A.~T. {Gallant},
  P.~{Hunt}, J.~{Kelly}, B.~{Kootte}, Y.~{Lan}, S.~F. {Paul}, M.~L. {Phan},
  M.~P. {Reiter}, B.~{Schultz}, D.~{Short}, J.~{Simonis}, C.~{Andreoiu},
  M.~{Brodeur}, I.~{Dillmann}, G.~{Gwinner}, J.~D. {Holt}, A.~A. {Kwiatkowski},
  K.~G. {Leach}, and J.~{Dilling}.
\newblock {Precision mass measurements of $^{125-127}$Cd isotopes and isomers
  approaching the $N=82$ closed shell}.
\newblock {\em ArXiv e-prints}, May 2017.

\bibitem{Zhang+2016}
Y.~H. {Zhang}, Y.~A. {Litvinov}, T.~{Uesaka}, and H.~S. {Xu}.
\newblock {Storage ring mass spectrometry for nuclear structure and
  astrophysics research}.
\newblock {\em Physica Scripta}, 91(7):073002, July 2016.

\bibitem{FRIB}
{https://groups.nscl.msu.edu/frib/rates/fribrates.html}.

\bibitem{Mumpower+12}
M.~R. {Mumpower}, G.~C. {McLaughlin}, and R.~{Surman}.
\newblock {Formation of the rare-earth peak: Gaining insight into late-time
  r-process dynamics}.
\newblock {\em \prc}, 85(4):045801, April 2012.

\bibitem{Neufcourt+2018}
L.~{Neufcourt}, Y.~{Cao}, W.~{Nazarewicz}, and F.~{Viens}.
\newblock {Bayesian approach to model-based extrapolation of nuclear
  observables}.
\newblock {\em Phys.\ Rev.\ C}, 98(3):034318, September 2018.

\bibitem{Giuliani+2018}
S.~A. {Giuliani}, G.~{Mart{\'{\i}}nez-Pinedo}, and L.~M. {Robledo}.
\newblock {Fission properties of superheavy nuclei for r -process
  calculations}.
\newblock {\em \prc}, 97(3):034323, March 2018.

\bibitem{Bulgac+18}
A.~{Bulgac}, S.~{Jin}, K.~{Roche}, N.~{Schunck}, and I.~{Stetcu}.
\newblock {Fission Dynamics}.
\newblock {\em arXiv e-prints}, June 2018.

\end{thebibliography}

\end{document}